\def\r{\rho}
\def\beg{\begin{equation}}
\def\eeq{\end{equation}}
\documentstyle[12pt]{article}
\begin{document}
\begin{center}
{\Large{\bf Polarization of the half-filled Landau level in quantum Hall
effect.}}
\vskip0.5cm
{\bf Keshav N. Shrivastava}\\
{\it School of Physics, University of Hyderabad,\\
Hyderabad  500 046, India.}
\end{center}
\vskip0.5cm
We develop a theory of the half-filled Landau level which shows that it is
a two-component fluid. One of the components has spin up and the other
spin down so that an electron spin resonance can occur. The 1/3 filled
level is fully spin polarized one-component fluid. We calculate the spin
polarization at the $\nu=1/2$ as a function of temperature and compare the
calculated result with the experimentally measured value. Our theory is in
very good agreement with the experimental data.
\vskip5.0cm
keshav@mailaps.org   Fax. +91-40-3010145
\newpage

\noindent{\bf 1.~Introduction.}

According to the classical Hall effect the transverse resistivity is
linearly proportional to the applied magnetic field and the slope of the
resistivity versus magnetic field can be used to measure the concentration
of electrons,
\beg
\r_{xy} = {B_z\over nec}\,\,\,,
\eeq
where $n$ is the electron concentration and $B_z$ is the magnetic field
along $z$ direction. In 1980 von Klitzing, Dorda and Pepper [2] reported
that instead of the linear dependence as shown above, a plateau occurs in
$\r_{xy}$ which can be measured so accurately that the value of the
$e^2/h$ can be deduced. Therefore, at the plateau, we have
\beg
{B\over nec} = {h\over ie^2}
\eeq
and hence, 
\beg
i = {n\over(eB/hc)}
\eeq
is an integer. We write the electron density as $n=n_o/A$, i.e. the number
of electrons per unit area so that the above equation becomes,
\beg
i = {n_o\over A(eB/hc)}
\eeq
which may be arranged as
\beg
B.A = {n_o\over i}\left({hc\over e}\right)
\eeq
which is another way of saying that there is a flux quantization,
$\phi_o=hc/e$ and $n_o$ and $i$ are integers. Therefore, von Klitzing et
al have detected flux quantization while measuring the Hall effect. Very
soon it was reported by Tsui, St\"ormer and Gossard [2] that not only $i=1$
but fractional quantization occurs. The value of $i=1/3$ was detected and
many other fractions 2/3, 2/5, 3/5 etc were found [3]. Laughlin [4,5] made
the theoretical efforts to find the wave function of a fractional charge
by introducing incompressibility. In the flux quantization condition 1/3
charge is equivalent to three times the flux
\beg
{hc\over{1\over3}e} = {3hc\over e}
\eeq
or the one third area,
\beg
B. {A\over 3} = n_o {hc\over e}\,\,\,.
\eeq
Similarly, if we consider the charge per unit area, then 1/3 charge per
unit area is equivalent to having one charge in three times the area
\beg
{{1\over3}e\over A} = {e\over3A}\,\,.
\eeq
The cause of incompressibility is not given in Laughlin's paper
and the force which holds the distances between particles constant
 is not mentioned. The charge can be 1/3 if area is $A$
or alternatively, if we give up incompressibility, the charge can be 1 in
thrice the area, $3A$. Laughlin's wave functions are defined in the
complex space in two dimensions and can not be used to determine those
properties which are often measured in a variety of experiments. There are
other theories in which it is suggested that either even number or odd
number of fluxes are attached to the electron to form a ``composite
fermion''. The odd number of fluxes such as one flux plus one electron
form a boson which obeys the Bose-Einstein statistics and hence can make a
Bose condensate. The even number of fluxes attached to the electron, form
fermions. A mixture of bosons and fermions need not always make
quasiparticles of mixed statistics but it has been reported that flux
tubes attached to electrons will have intermediate statistics. At this
time these theories are not supported by experimental evidence and
attachment of flux tubes to electrons is not confirmed.

We have found [6] that a simple theory explains the experimental data
without contradicting the ideas of fractional charge. The fractions which
we tabulated in 1985 are in full agreement with Fig. 18 of St\"ormer's 
Nobel lecture [7].
We have examined considerable amount of data in which fractions are
experimentally observed and in each and every case, the observed values
are found to be the same as those predicted [8,9]. Some of the
experimentally observed masses are equal and this equality of masses is
well explained by the particle-hole symmetry which in our theory is
related to the Kramers conjugate states [10]. The high Landau levels also
fit very well in our theory [11]. The magnetic moment of the electron is
slightly changed due to an effective spin-orbit interaction [12]. It is
found that the rate of sweep of magnetic field plays an important role in
determining whether the transverse resistivity should be zero or finite.
The field is detected by nuclear-magnetic resonance. There is obviously a
phase factor problem which determines whether the system prefers a zero or
a finite value of $\rho_{xx}$ [13]. There is a phase transition and
Goldstone modes occur due to flux motion, in bilayers [14]. Dementyev et
al [16] have obtained the electron-spin polarization by measuring the
$^{71}$Ga nuclear magnetic resonance in 300 \AA \, wide GaAs wells
separated by 3600 \AA \, wide Al$_{0.1}$Ga$_{0.9}$As barriers at a magnetic
field which corresponds to resistivity plateaus at $\nu=1$ and 1/3 where
$\r_{xy}=h/\nu e^2$ in the quantum Hall effect. These are very unique NMR
measurements which require a suitable theory [15]. The electron-spin
polarization is obtained from the ratio of Knight shifts of the NMR lines as
\beg
P(\nu={1\over2}, T) = {K_s(\nu={1\over2},T)f({1\over2})\over
K_s({1\over3})f({1\over3})}
\eeq
where $K_s(\nu)$ are the Knight shifts for the fields at which
$\nu=1/2$ which is partially polarized and $\nu=1/3$ which is fully
polarized and $f(\nu)=2n/[\omega\r(o)]$ with $n$ as the electron density
and $\omega$ the well width. Here $\rho(o)$ is the 3-dimensional electron
density at the center of the quantum well. There is an orientation angle
$\theta$ between the total magnetic field and the growth axis of the
sample. 

In this paper, we use our theory to understand the polarization at 1/3 and
show that there is a two-component fluid at $\nu={1\over2}$ with one
component having spin up and the other down. Therefore, there is an
electron spin resonance energy between the up and the down components. We
calculate the electron spin polarization at the $\nu={1\over2}$ and
compare it with that experimentally measured by using the NMR at the
quantum Hall effect fields at $\nu = {1\over2}$ and $\nu = {1\over3}$. We find
that the predicted temperature dependence is in accord with that measured
by Dementyev et al [16].

\noindent{\bf 2. Theory.}

Polarization at 1/3. According to our theory first given in ref. 6 and
later on in refs. [8-15], the quantum Hall plateaus occur at $\nu$ given
by two series,
\beg
\nu_\pm = {l+({1\over2})-s\over2l+1}.
\eeq
For $s={1\over2}$, we obtain
\beg
\nu_+ = {l\over2l+1}
\eeq
and for $s=-1/2$, we get
\beg
\nu_-={l+1\over2l+1}\,\,\,.
\eeq
These two series are sufficient to explain all of the relevant fractions
observed in the experimental data. We take the limit of $l\to\infty$,
which gives $\nu_+ = {1\over2}$. Similarly, the $\nu_-$ series at
$l\to\infty$ also gives $\nu_-={1\over2}$. Therefore at $\nu={1\over2}$,
the value approaches from both the right hand side of 1/2 as well as from
the left hand side so that the liquid has two components. We call
$\nu_+=1/2$ as $A^{(+)}$ and $\nu_-={1\over2}$ as $B^{(-)}$. The values
with + sign always have spin +1/2 and the values $\nu_-$ always have spin
-1/2. For various values of $l$ $\nu_\pm$ series give,
\begin{equation}
\begin{array}{ccccccccc}
l & = & 0 & 1 & 2 & 3 & 4 & 5 &\infty\\
\\
\nu_+ & = & 0 & {1\over3} & {2\over5} & {3\over7} & {4\over9} & {5\over11} &
{1\over2}\\
\\
\nu_- & = & 1 & {2\over3} & {3\over5} & {4\over7} & {5\over9} & {6\over11}&
{1\over2}\end{array}
\end{equation}
Therefore $\nu={1\over3}$ is fully spin polarized with polarization $P=1$.
Howerver, there will be entropy and therfore, there is always a
possibility that some of the spins are not aligned to the value of +1/2
otherwise the state with $\nu=+1/3$ is fully polarized. Thus we fix the
polarization of various states as given above. Any deviation may arise
due to thermodynamics. Since $\nu=1/2$ arises from both +1/2 as well as
-1/2, these states are separated in energy by that of the electron spin
resonance, 
\beg
k_BT_z = g^*\mu_BB_{tot}
\eeq
where $g^*$ is the usual Lande's splitting factor, $g^*=2.0023$ for free
electrons and $g^*=-0.44$ in the semiconductor GaAs. If the energy of the
$N$ electrons is required, we can multiply the above by $N$. For one
electron in the $s=+1/2$ state the energy is ${1\over2}g^*\mu_BB_{tot}$ 
and in the $s=-1/2$ state it is $-{1\over2}g^*\mu_BB_{\rm tot}$. In
the experiment of Dementyev et al, $T_z\simeq 1.63 K$ is a temperature
which defines the energy difference between the two spin states. Usually,
three components of the triplet and one singlet should arise but in our
formula (13) only two states are found at $\nu_\pm=1/2$. Thus we have the
 fixed value of polarization at $\nu=1/3$
and a two component liquid state at $\nu_\pm=1/2$. We will now calculate
the polarization at $\nu_\pm=1/2$ as per usual simple Maxwell-Boltzman
distribution. We consider only a two-level system. Since the particles in
these levels are separated by energy given by eq. (14), the population in
the two levels will be,
$$
{N_1\over N} = {\exp({1\over2}g^*\mu_BB/k_BT)\over
\exp({1\over2}g^*\mu_BB/k_BT) + \exp(-{1\over2}g^*\mu_BB/k_BT)}
$$
and
\beg
{N_2\over N} = {\exp(-{1\over2}g^*\mu_BB/k_BT)\over
\exp({1\over2}g^*\mu_BB/k_BT) + \exp(-{1\over2}g^*\mu_BB/k_BT)}.
\eeq
Here $N_1$ and $N_2$ are the populations of the lower and upper levels for
a positive value of $g^*$ and $N = N_1 + N_2$  is the total number of
electrons. In case $g^*$ is negative, the levels get interchanged. The
projection of the magnetic moment of the upper state along the field
direction is $-{1\over2}g^*\mu_B$ and that of the lower state is
$+{1\over2}g^*\mu_B$ which again get interchanged if the sign of $g^*$ is
negative. We make the change of variables as,
\beg
x = {1\over2} g^*\mu_BB/k_BT\,\,\,.
\eeq
The magnetization of $N$ sites per unit volume is,
\beg
M = (N_1-N_2)\mu_B({1\over2}g^*) = {1\over2}Ng^*\mu_B {e^x-e^{-x}\over
e^x+e^{-x}} = {1\over2} Ng^*\mu_B \tanh x.
\eeq
The high-temperature expansion of $\tanh x = x$ may be found for $x\ll 1$ so
that,
\beg
M=N({1\over2}g^*\mu_B)^2B/k_BT.
\eeq
Thus the magnetization and hence the polarization in the state which has a
population of $N_1$ varies as the inverse temperature. From this result
the susceptibility also varies as $1/T$. At $\nu=1/2$, the quantum Hall
system behaves like a two level system. Near the peak there is a small
region which requires the wave vector dependent response given in ref.
[8]. 

\noindent{\bf 3. Comparison with experimental data.}

Dementyev et al [16] have performed the experimental measurements of the
electron spin polarization which we compare with our theory. Two
cases have been measured. Case I has forty wells of width 300 \AA
separated by 3600 \AA wide Al$_{0.1}$Ga$_{0.9}$As barriers. The angle
between the sample growth axis and the direction of the total magnetic
field is $\theta38.4^o$, $B_{tot}=7.03$T. The case II has $\theta=0.0^o$
and $B_{tot}=5.52$T. The spin polarization at $\nu=1/2$ for both the cases
is given as a function of temperature with $T_z=2.08$K in case I and
$T_z=1.63$K in case II.  In both the cases the polarization varies as
$1/T$ as predicted by our theory for $T>0.6$K. There is no analog of the
ferromagnetic phase and if any, its transition temperature will be
$T_c = 0.0K$. Therefore, there is no region where we can see the effect of
the dependence on the wave vector of the modes. For temperatures less than
0.6 K there may be a contribution to the polarization which depends on the
wave vector of the response function. However, we know that a Goldstone
boson emerges at low temperatures in the two component fluid so that the
polarization will obey a critical exponent characteristic of a phase
transition. The polarization at $\nu=1/2$ measured by Dementyev et al is
shown in Fig.1  as a function of temperature. Two calculated curves are
also plotted alongwith the data. It is clear that the data agrees with the
predicted behaviour at most of the temperatures, $0.6<T<3.4$K. It is
possible that there is a region below 0.6 K where wave vector dependence
should be considered. An effort is made to compare the polarization as a
function of $(1-T/T_c)^\alpha$ in search of an exponent below $T=1$K. It
is found that for $T<1$K, $\alpha=0.15$ as shown in the ln-ln plot shown
in Fig. 2 for $0.26 <T<0.76$ K. Apparently, the Knight shifts used to
measure the polarization provide this value for the electron spin
polarization but such a small value for the exponent of polarization is
not expected from the scaling theory. Similar study of one more sample
has been reported by Melinte et al [17]. We have extracted some
representative points from their measurements and shown them in Fig. 3
alongwith the curve calculated by us. Apparently, our theory is in
reasonable agreement with the data. 

\noindent{\bf 4. Conclusions.}

We conclude that there is a two component fluid at $\nu=1/2$ with one
component having spin +1/2 and the other component having spin -1/2. The
polarization near or above a temperature of 1 K varies as the inverse
temperature. In this respect our theory agrees with the experimental data.
It should be noted that there is no evidence of attaching even or odd flux
quanta to an electron and there is no evidence for the existence of any
composite fermions or composite bosons. Assuming that flux was attached to
the electrons, the field $B$ in (18) will be replaced by,
\beg
 {B^*}={B-2p\r\phi_{o}}
\eeq
where $p$ is an integer so that $2p$ is an even number, $\r$ is the charge
 density per unit area and $\phi_{o}$ is the unit flux. When we put $p$ =
 0, 1, 2, 3, ..., etc., we get different values of $B^*$ and hence from
 (18) different values of the magnetization. Therefore, magnetization 
should show plateaus for different values of $p$. In the present problem,
the polarization as a function of temperature should show plateaus. No 
such plateaus are found and hence there is no evidence of attachment of
fluxex. If spin is completely free, in the half filled Landau level a 
four-component fluid with a singlet and three components of the triplet
should occur. Such a fluid is not observed in the experiments. 
In a previous eprint [15] we have shown that Jain's theory
is internally inconsistent. From the present discussions it is clear
 that the fluxes are not attached to the electrons and four components
are not found. Hence Jain's theory [18] of composite fermions is not
 applicable to the experimental measurements.
\newpage
\noindent{\bf References.}
\begin{enumerate}
\item K. von Klitzing, G. Dorda and M. Pepper, Phys. Rev. Lett. {\bf45},
 494 (1980).
\item D.C. Tsui, H.L. St\"ormer and A.C. Gossard,Phys. Rev. Lett. {\bf48},
 1559 (1982).
\item J.R. Eisenstein and H.L. St\"ormer, Science, {\bf248}, 1510 (1990).
\item R.B. Laughlin, Phys. Rev. B{\bf27}, 3383 (1983).
\item R.B. Laughlin, Phys. Rev. Lett. {\bf50}, 1395 (1983).
\item K.N. Shrivastava, Phys. Lett. A{\bf113}, 435 (1986).
\item H.L. St\"ormer, Rev. Mod. Phys. {\bf71}, 875 (1999).
\item K.N. Shrivastava, CERN SCAN-0103007.
\item K.N. Shrivastava, in Frontiers in Physics, edited by B.G. Sidharth,
Kluwer Press, 2001.
\item K.N. Shrivastava, Mod. Phys. Lett. {\bf13}, 1087 (1999).
\item K.N. Shrivastava, Mod. Phys. Lett. {\bf14}, 1009 (2000); cond-mat/0103604.
\item K.N. Shrivastava, cond/mat 0104004.
\item K.N. Shrivastava, cond/mat 0104577.
\item K.N. Shrivastava (in preparation).
\item K.N. Shrivastava, cond-mat/0105559.
\item A.E. Dementyev, N.N. Kuzma, P. Khandelwal, S.E. Barrett, L.N.
      Pfeiffer and K.W. West, Phys. Rev. Lett. {\bf83}, 5074 (1999).
\item S. Melinte, N. Freytag, M. Horvatic, C. Berthier, L.P. Levy, V.
      Bayot and M. Shayegan, Phys. Rev. Lett. {\bf84}, 354 (2000).
\item J. K. Jain, Phys. Rev. Lett. {\bf63}, 199 (1989).
\end{enumerate}
\end{document}